\documentclass[longauth]{aa}

\usepackage{graphicx}
\usepackage[dvipsnames]{xcolor}
\usepackage{txfonts}
\usepackage[colorlinks=true, citecolor=blue]{hyperref}

\usepackage{multirow}
\usepackage{booktabs}
\usepackage{pdflscape}
\usepackage{placeins}
\usepackage{caption}

\newcommand{\hb}{H$\beta$}
\newcommand{\ha}{H$\alpha$}
\newcommand{\hg}{H$\gamma$}
\newcommand{\hd}{H$\delta$}

\newcommand{\oiii}{[O\,{\footnotesize III}]}

\newcommand{\nii}{[N\,{\footnotesize II}]}

\newcommand{\kms}{km s$^{-1}$}
\newcommand{\es}{erg s$^{-1}$}
\newcommand{\myr}{$M_{\odot}$ yr$^{-1}$}
\newcommand{\msun}{$M_{\odot}$}

\newcommand{\mstar}{$M_{\star}$}

\newcommand{\mbh}{$M_{\rm BH}$}

\newcommand{\lbol}{$L_{\rm bol}$}

\newcommand{\hst}{HST}

\newcommand{\sigmastar}{$\sigma_\star$}

\newcommand{\dn}{$D_n4000$}
\newcommand{\nev}{[NeV]}
\newcommand{\lnev}{$L_{\rm [NeV]}$}
\newcommand{\loiii}{$L_{\rm [OIII]}$}
\newcommand{\eddratio}{$\lambda_{\rm Edd}$}

\newcommand{\mistral}{\textsc{Mistral}}
\newcommand{\blackdawn}{{\sc Black Dawn}}
\newcommand{\shark}{\textsc{Shark}}
\newcommand{\gaea}{\textsc{Gaea}}
\newcommand{\tng}{\textsc{TNG}}

\usepackage{CJKutf8} %package for Chinese characters
\newcommand{\zh}[1]{\begin{CJK}{UTF8}{bsmi}#1\end{CJK}}

\begin{document} 

   \title{High-ionization coronal lines trace quasar-like activity in recently quenched galaxies at high redshift}

  \titlerunning{High-ionization coronal lines trace radiatively efficient AGN in recently quenched galaxies at high redshift}
  \authorrunning{F. Valentino et al.}
\author{F. Valentino\inst{1,2} \and  
           K. Ito\inst{1,2} \and
           M. Farcy\inst{3} \and
           F. Fontanot\inst{4,5} \and
           C. Lagos\inst{6,7,1} \and
           G. De Lucia\inst{4,5} \and
           M. Hirschmann\inst{3,4} \and
           G. Brammer\inst{1,8} \and
           V. Kokorev\inst{9} \and
           M. Hamadouche\inst{10} \and
           P. Zhu (\zh{朱芃佩})\inst{1,2,11} \and
           G. Scarpe\inst{1,2} \and
           A. Pensabene\inst{1,2} \and
           K. E. Whitaker\inst{10} \and
           W. M. Baker\inst{12} \and
           P. Araya-Araya\inst{1,2} \and
           J. Antwi-Danso\inst{13,14,10} \and
           D. Ceverino\inst{15,16} \and
           A. L. Faisst\inst{17} \and
           S. Fujimoto\inst{13,14} \and
           S. Gillman\inst{1,2} \and
           O. Ilbert\inst{18} \and
           C. K. Jespersen\inst{19} \and
           T. Kakimoto\inst{20,21} \and
           M. Kubo\inst{22} \and
           A. W. S. Man\inst{23} \and
           G. Magdis\inst{1,2} \and
           M. Onodera\inst{24,21,20} \and
           R. Shimakawa\inst{25} \and
           M. Tanaka\inst{20,21} \and
           S. Toft\inst{1,8} \and
           L. Xie\inst{26,21} \and
           J. R. Weaver\inst{27} \and
           P.-F. Wu\inst{28}
          }
   \institute{
   Cosmic Dawn Center (DAWN), Denmark
   \and DTU Space, Technical University of Denmark, Elektrovej 327, DK-2800 Kgs. Lyngby, Denmark
   \and \'Ecole Polytechnique F\'ed\'erale de Lausanne (EPFL), Observatoire de Sauverny, Chemin Pegasi 51, CH-1290 Versoix, Switzerland
   \and INAF -- Astronomical Observatory of Trieste, Via G. B. Tiepolo 11, 34143 Trieste, Italy
   \and IFPU -- Institute for Fundamental Physics of the Universe, Via Beirut 2, 34151 Trieste, Italy
   \and International Centre for Radio Astronomy Research (ICRAR), M468, University of Western Australia, 35 Stirling Hwy, Crawley, WA 6009, Australia
   \and ARC Centre of Excellence for All Sky Astrophysics in 3 Dimensions (ASTRO 3D)
   \and Niels Bohr Institute, University of Copenhagen, Jagtvej 128, 2200 Copenhagen N, Denmark
   \and Department of Astronomy, The University of Texas at Austin, Austin, TX 78712, USA
   \and Department of Astronomy, University of Massachusetts, Amherst, MA 01003, USA
   \and INAF-Osservatorio Astrofisico di Arcetri, Largo Enrico Fermi 5, I-50125 Firenze, Italy
   \and DARK, Niels Bohr Institute, University of Copenhagen, Jagtvej 155A, DK-2200 Copenhagen, Denmark
   \and David A. Dunlap Department of Astronomy and Astrophysics, University of Toronto, 50 St. George Street, Toronto, Ontario, M5S 3H4, Canada
   \and Dunlap Institute for Astronomy and Astrophysics, 50 St. George Street, Toronto, Ontario, M5S 3H4, Canada
   \and Departamento de Fisica Teorica, Modulo 8, Facultad de Ciencias, Universidad Autonoma de Madrid, 28049 Madrid, Spain
   \and CIAFF, Facultad de Ciencias, Universidad Autonoma de Madrid, 28049 Madrid, Spain
   \and IPAC, California Institute of Technology, 1200 E. California Blvd. Pasadena, CA 91125, USA
   \and Aix Marseille Univ, CNRS, LAM, Laboratoire d'Astrophysique de Marseille, 13388 Marseille CEDEX 13, France
   \and Department of Astrophysical Sciences, Princeton University, Princeton, NJ 08544, USA
   \and Department of Astronomical Science, The Graduate University for Advanced Studies, SOKENDAI, 2-21-1 Osawa, Mitaka, Tokyo 181-8588, Japan
   \and National Astronomical Observatory of Japan, 2-21-1 Osawa, Mitaka, Tokyo 181-8588, Japan
   \and School of Science, Kwansei Gakuin University, Sanda, Hyogo 669-1337, Japan
   \and Department of Physics \& Astronomy, University of British Columbia, 6224 Agricultural Road, Vancouver, BC V6T 1Z1, Canada
   \and Subaru Telescope, National Astronomical Observatory of Japan, National Institutes of Natural Sciences (NINS), 650 North A'ohoku Place, Hilo, HI 96720, USA
   \and Waseda Institute for Advanced Study (WIAS), Waseda University, 1-21-1 Nishi-Waseda, Shinjuku, Tokyo 169-0051, Japan
   \and Astrophysics Center, Tianjin Normal University, Binshuixidao 393, 300387, Tianjin, China
   \and MIT Kavli Institute for Astrophysics and Space Research, 70 Vassar Street, Cambridge, MA 02139, USA
   \and Graduate Institute of Astrophysics, National Taiwan University, Taipei 10617, Taiwan
   }
   \date{Received --; accepted --}
    \abstract{We report the detection of the high-ionization line \nev$\lambda$3427 in the \textit{JWST}/NIRSpec archival spectra of 6 massive quenched galaxies at $z \sim 1.5$--$4.5$, identified from a parent sample of 87 systems. With an ionization potential of approximately 97 eV, \nev\ can only be produced by strong nuclear activity in these massive systems, providing a clean and unambiguous tracer of highly accreting supermassive black holes uncontaminated by residual star formation. For 4 of the 6 \nev-detected systems, we detect broad \ha\ emission ($\mathrm{FWHM} \gtrsim 4000$ \kms), yielding black hole masses of $M_{\rm BH} = 10^{8.5\text{--}9.5}\,M_\odot$, consistent with local scaling relations with stellar mass and velocity dispersion. The \nev\ luminosities imply quasar-like bolometric outputs ($L_{\rm bol} = 10^{45\text{--}46}$ \es) and Eddington ratios of $\lambda_{\rm Edd} \approx 10$--$50$\%, with black hole accretion rates of a few \myr\ that match or exceed the residual star formation rates in the most extreme cases. The strongest \nev\ emitters are preferentially found in the youngest post-starburst systems ($D_n4000 \lesssim 1.3$), while old quenched galaxies are systematically devoid of such activity, a trend independently reproduced by theoretical models. These results reveal that intense, radiatively efficient SMBH growth can persist several hundred Myr after the main quenching epoch, with duty cycles of approximately 100--200 Myr. They also underscore the importance of very high accretion episodes and rates in the theoretical models that seek to reproduce the earliest quenched galaxies in the universe.}

  \maketitle
  \nolinenumbers

\section{Introduction}
Galaxy quenching in the early Universe remains one of the central open questions in galaxy evolution. Massive quiescent galaxies are already in place at $z>3$ (e.g., \citealt{schreiber_2018c, valentino_2020a, valentino_2023, carnall_2022, nanayakkara_2024, baker_2025, baker_2025_dja, merlin_2025, yang_2025_miri, antwi-danso_2025, stevenson_2026}), implying that star formation must proceed rapidly and efficiently before being suppressed
for extended periods of time at even earlier epochs \citep{carnall_2023b, carnall_2024, glazebrook_2024, deGraaff_2025, weibel_2025}. At the same time, luminous quasars powered by billion-solar-mass black holes are observed at $z>6$ (e.g., \citealt{fan_2006, banados_2018, fujimoto_2022, yang_2023_aspire}), demonstrating that supermassive black holes (SMBHs) undergo episodes of extremely rapid growth within the first few Gyr of cosmic time. Understanding how these two phenomena -- the early progression and shutdown of star formation and the rapid assembly of SMBHs -- are connected is a fundamental challenge for both galaxy evolution and black hole studies.

Energetic feedback from accreting SMBHs is widely invoked as the mechanism capable of reconciling these observations. In semi-analytical models and cosmological hydrodynamical simulations, feedback from active galactic nuclei (AGN) is required to reproduce the high-mass end of the stellar mass function and the existence of passive massive galaxies (e.g., \citealt{croton_2006, somerville-dave_2015, naab-ostriker_2017, alexander_2025, farcy_2025, kimmig_2025, delucia_2024, lagos_2024, chandro-gomez_2025, chaikin_2026, araya-araya_2026}). However, the effectiveness of this mechanism depends critically on when black holes accrete, for how long, at which luminosities, and how they impact their surrounding environment. Tracking the SMBH growth before, during, or after the quenching process, possibly interspersed with episodes of rejuvenation \citep{fontanot_2025b, delucia_2026}, remains observationally unconstrained and complicated due to the timescales involved. Establishing the relative timing between star formation suppression and luminous AGN phases, at least at the population level, is therefore essential.

In theoretical frameworks, the timing, duty cycle, and strength of AGN episodes are fundamental outputs of sub-grid physical recipes. Different implementations of black hole accretion and feedback -- for example in their dependence on gas supply or halo properties, their initial seeding, and freedom to accrete and grow at capped or unconstrained rates --- can lead to markedly different quenching timescales and number densities of massive passive galaxies at high redshift (\citealt{lagos_2025} for a summary). While many models are calibrated to reproduce local observables, they often diverge in their predictions for the formation pathways of the earliest quenched systems. Small changes in feedback prescriptions and their free parameters can drastically alter the abundance, ages, and structural properties of massive galaxies \citep{genel_2019, fontanot_2020, delucia_2024, lagos_2025, chandro-gomez_2025}. Constraining when and how efficiently SMBHs accrete around the quenching phase therefore provides a powerful empirical lever arm on the physical assumptions encoded in theoretical models.

From the black hole perspective, a complementary question emerges: can substantial SMBH growth occur in galaxies that have already largely ceased forming stars? Most observational studies start from AGN-selected samples, which include systems that are still actively forming stars. In contrast, distant massive quenched galaxies offer a distinct and underexplored environment for studying SMBH growth. Despite well-known modeling systematics, their relatively simple stellar populations allow more robust measurements of stellar masses and velocity dispersions compared to star-forming systems, where complex star formation histories, nebular emission, the absence of strong stellar absorption features, and dust introduce additional uncertainties. As a result, quenched galaxies provide a cleaner laboratory to test black hole scaling relations and to assess whether rapidly accreting SMBHs are consistent with, or offset from, local co-evolutionary relations with stellar masses or velocity dispersions \citep{kormendy_2013,greene_2020}.

Identifying AGN activity in these systems requires diagnostics insensitive to residual star formation, as classical line-ratio diagnostics such as the BPT diagram \citep{baldwin_1981} can be contaminated by it. The high-ionization \nev$\lambda3427$
 coronal line offers a particularly clean alternative: with an ionization potential of $97.11$ eV, it cannot be produced by stellar processes in metal-rich, evolved systems, making it a robust tracer of strong nuclear activity in massive galaxies \citep{gilli_2010, mignoli_2013, cleri_2023, vergani_2018, vergani_2025, trakhtenbrot_2025}. This fine-structure line is thought to arise in an intermediate region between the narrow and broad line regions, and is less affected by dust than other coronal lines \citep{rodriguez-ardila_2025}. Current photoionization models can account for its luminosity and possible extension on kiloparsec scales \citep{mckaig_2024}, although extreme AGN-driven shocks and outflows can contribute to part of the emission \citep{leung_2021}. At $z\gtrsim0.8$, \nev\ falls within the \textit{JWST}/NIRSpec wavelength range, enabling systematic searches for powerful AGN in the rest-frame near-UV of quenched systems.

Recent \nev\ studies have focused on AGN-selected samples at low and intermediate redshifts \citep[e.g.,][]{doan_2025}, and have identified systems consistent with galaxies immediately before quenching \citep{barchiesi_2024, barchiesi_2025}. However, ground-based spectra are often insufficient to detect the underlying stellar populations of individual sources, especially with increasing redshift. Here we adopt a complementary, galaxy-first approach, more akin to that in \citet{vergani_2018}: we select massive quenched galaxies in the literature and search for signatures of ongoing, radiatively efficient SMBH accretion within them. This allows us to directly test whether luminous, quasar-like accretion episodes can occur after quenching.

In this work, we report the detection of bright \nev\ emission in an archival sample of massive quenched galaxies at $z\sim1-4.5$ observed with \textit{JWST}/NIRSpec. The inferred bolometric luminosities reach values consistent with those of low luminosity quasars at similar redshifts ($L_{\rm bol}=10^{45-46}$ \es), revealing episodes of intense and radiatively efficient black hole growth several hundred Myr after the primary shutdown of star formation. These observations simultaneously provide new constraints on the timing and duty cycle of AGN activity in the quenching sequence and offer direct insight into how rapidly growing SMBHs assemble in already massive, evolved host galaxies in the early Universe. Throughout this work, we adopt a $\Lambda$CDM cosmology with $\Omega_{\rm m} = 0.3$, $\Omega_{\Lambda} = 0.7$, and $H_0 = 70\,\mathrm{km\,s^{-1}\,Mpc^{-1}}$.

% Figure: Stellar mass, Dn4000$
\begin{figure}
    \centering
    \includegraphics[width=\columnwidth]{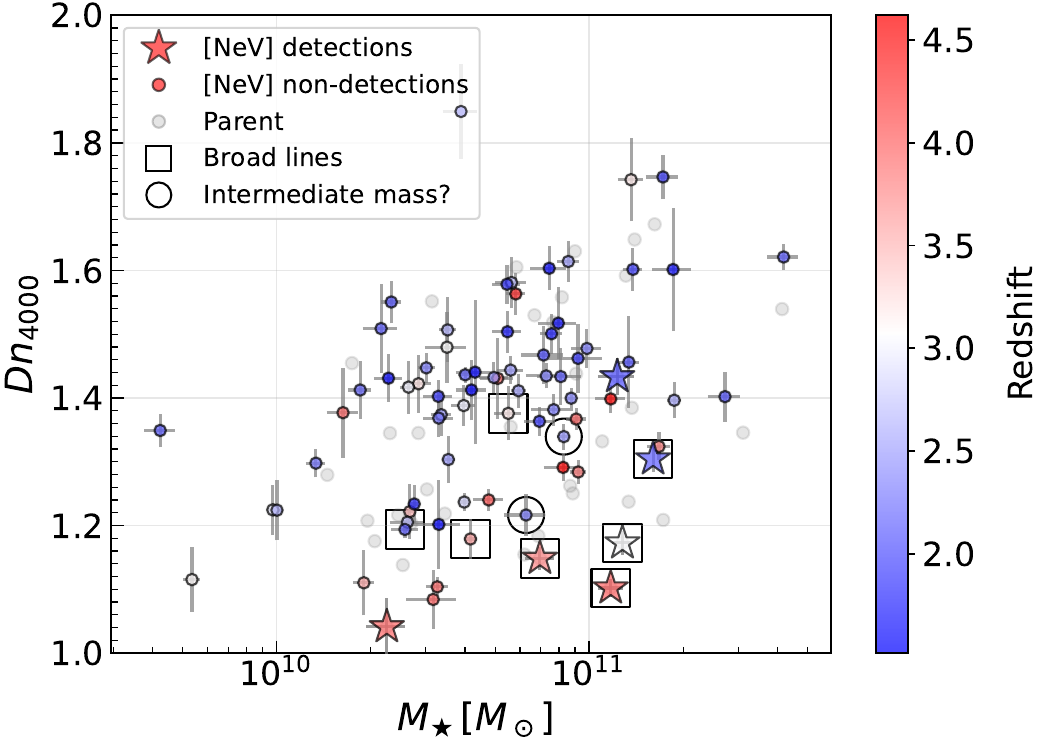}
    \caption{Stellar mass and \dn\ distribution of our targets. The sample with \nev\ coverage is colored according to redshift. Filled stars and circles indicate \nev\ detections and upper limits, respectively. The rest of the parent sample is marked with gray circles. Broad \ha\ line detections ($\mathrm{FWHM}\gtrsim 4000$ \kms) and possible intermediate black holes ($\mathrm{FWHM}\approx 2000$ \kms) are indicated with open black squares and circles.}
    \label{fig:m_dn}
\end{figure}

\section{Sample selection and methods}
Our parent sample consists of the 149 quiescent galaxies at $z\sim1-4.5$ compiled by \cite{ito_2025_deepdive}, selected on the basis of their rest-frame $UVJ$ colors, a specific star formation rate (sSFR) threshold $10\times$ below the main sequence, and/or the strength of the 4000 \AA\ break as parameterized by the \dn\ index \citep{balogh_1999}. The sample was selected from the DAWN JWST Archive (DJA), with medium or high-resolution NIRSpec/MSA spectra (v4) reduced with {\sc msaexp} \citep{degraaff_2024_rubies, heintz_2025, valentino_2025} and flux-calibrated to match the available NIRCam photometry \citep{valentino_2023}. We refer the reader to \cite{ito_2025_deepdive} for details on the sample selection and data reduction.

% Figure: Spectra
\begin{figure*}
    \centering
    \includegraphics[height=0.858\textheight]{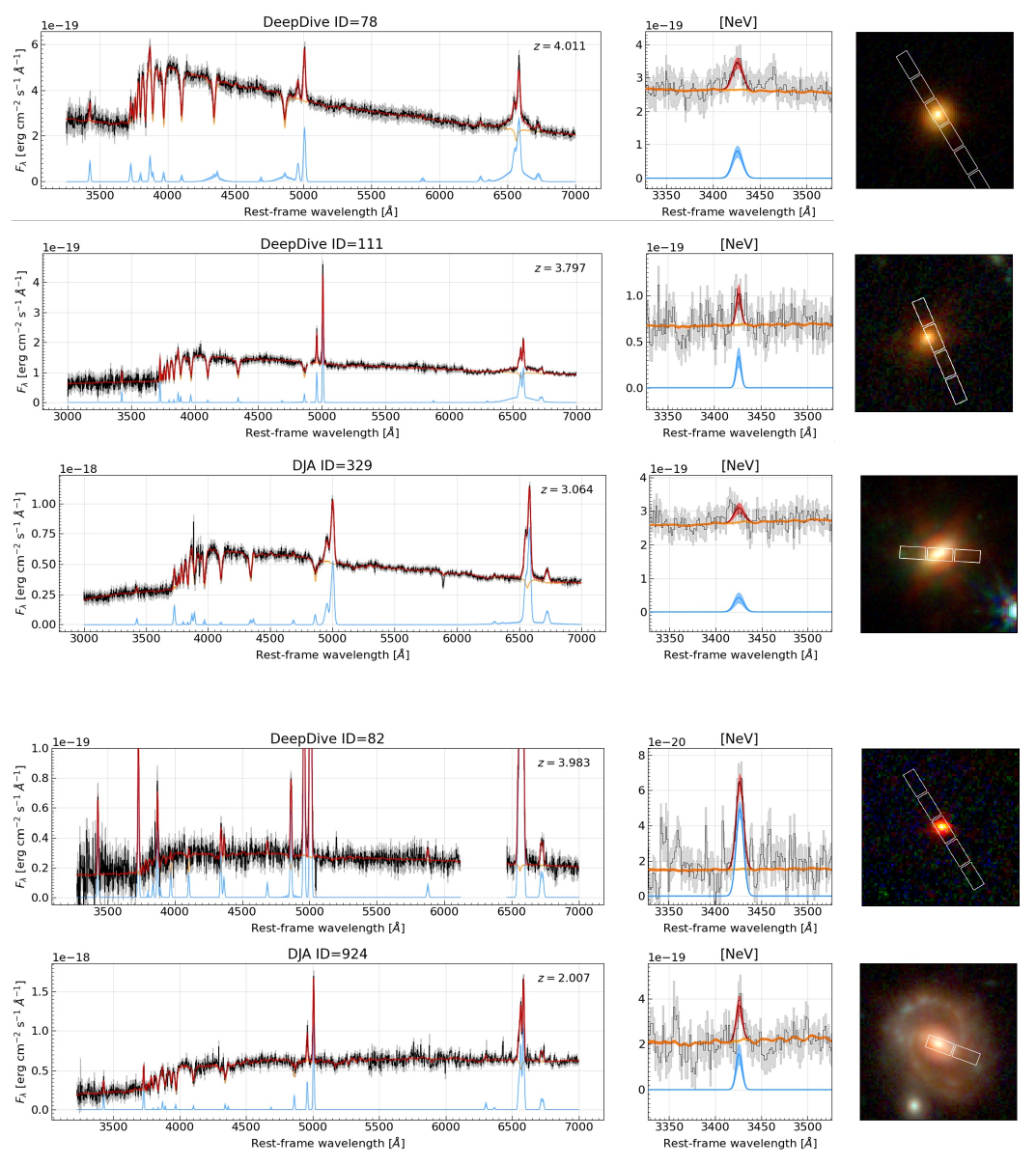}
    \caption{NIRSpec spectra of sources with \nev\ detections. Black and gray lines indicate the observed spectra and their uncertainties, while the red line shows the best-fit model with {\sc pPXF} (the orange and blue lines show the stellar and gas templates separately). The central panels show a zoom-in on the region around \nev. The right panels show $3^{"} \times 3^{"}$ RGB images of the targets. The ID and redshift of each source are indicated in the left panels. The sixth \nev\ detection included in this analysis is presented in \cite{ito_2025_agn}.}
    \label{fig:spectra}
\end{figure*}

Of the 149 quiescent galaxies (QGs) in \cite{ito_2025_deepdive}, 86 (58\%) have spectral coverage of the \nev\ line. The subsample spans a similar range in redshift, stellar mass, and \dn\ as the parent sample, without introducing any significant sampling bias (Figure \ref{fig:m_dn}). Stellar masses and star formation rates (SFRs) averaged over 100 Myr were derived from photometric spectral energy distribution (SED) modeling using {\sc Bagpipes} \citep{carnall_2018}, using \cite{bruzual_2003} models, the \cite{kroupa_2001} IMF, a \cite{calzetti_2000} dust attenuation law, emission lines with a fixed ionization parameter $\log{U}=-3$ \citep{byler_2017}, and a double power law SFH (\citealt{ito_2025_deepdive}; see \citealt{hamadouche_2026} for alternative SFH prescriptions and the effect of metal abundances). With the exception of four sources, the sample is covered by public {\it Chandra} observations, including 10 individual detections \citep{ito_2025_deepdive}. For the 72 sources covered by {\it Chandra}, but undetected, we set an upper limit on the 2-10 keV rest--frame X-ray luminosity (uncorrected for obscuration) based on the limiting sensitivity in the observed full {\it Chandra} band in each field, assuming a power-law index $\Gamma=1.8$. To assess the AGN contribution across the SEDs, we tested physically motivated AGN templates with {\sc XCigale} \citep{yang_2020_xcigale}, including X-ray emission when available, following the setup in \cite{ito_2025_agn}. The AGN templates typically contribute $<20$\% of the emission at rest-frame optical wavelengths, and stellar masses derived with and without AGN templates agree to within $0.1$ dex. These results align with findings for the average quiescent galaxy population from stacking analyses \citep{ito_2022}, and the AGN contribution to the host emission is also comparable to the rest-frame optical emission of faint QSOs at high redshift \citep[][]{ding_2025}.
% Figure: [NeV], X-ray luminosities as a function of redshift
\begin{figure}
    \centering
    \includegraphics[width=\columnwidth]{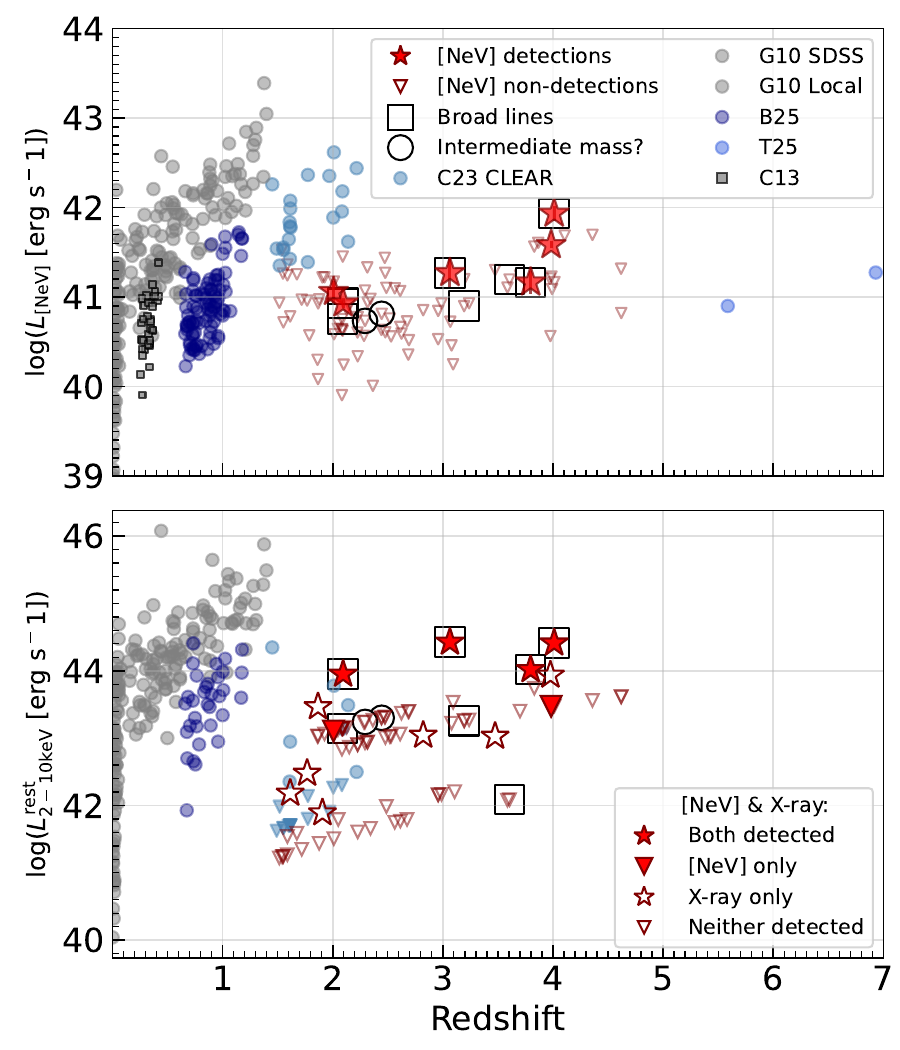}
    \caption{Redshift and luminosity distributions of our QG sample at $z \sim 1-4.5$ with \nev\ coverage (\lnev, top panel) and their X-ray counterparts where available ($L_{\rm 2-10\,keV}$, bottom panel). \nev\ detections and upper limits are marked with filled red stars and empty triangles in the top panel. In the bottom panel, filled symbols indicate \nev\ detections with X-ray coverage (filled stars and triangles indicate X-ray detections and upper limits, respectively). Empty symbols mark the rest of our \nev-undetected QG sample covered by \textit{Chandra} observations (empty red stars and triangles indicate X-ray detections and upper limits, respectively). In both panels, QGs with broad ($\mathrm{FWHM} \gtrsim 4000$ \kms) or intermediate ($\mathrm{FWHM} \approx 2000$ \kms) \ha\ emission are indicated with open black squares and circles, respectively. In both panels, gray circles indicate the sample of local and SDSS Seyferts in \cite{gilli_2010}, light blue circles mark the \hst/CLEAR star-forming sample at cosmic noon from \cite{cleri_2023}, and navy circles indicate the obscured AGN in \cite{barchiesi_2025}. In the top panel, two sources observed with NIRSpec at $z > 5$ \citep{trakhtenbrot_2025, chisholm_2024, scholtz_2025} are shown as cobalt circles, while dark gray squares indicate the ``post-starburst QSO'' sample in \citet{cales_2013}.}
    \label{fig:z_nev}
\end{figure}

To estimate the stellar velocity dispersions and emission line properties, we simultaneously modeled the stellar continuum and ionized gas emission, including \nev, in the observed spectra with pPXF \citep{cappellari_2017}. We used C3K stellar spectral libraries from the Flexible Stellar Population Synthesis (FSPS) models \citep{conroy_2010} and a \cite{kroupa_2001} IMF. We allowed for multiplicative polynomials up to the second degree and masked the regions affected by the detector gaps. The velocity centroid and dispersion of the stellar and gas components were left free to vary, but we tied the redshift and velocity dispersion of all emission lines. We then bootstrapped the modeling 500 times, sampling the noise spectrum, to obtain more conservative uncertainties. In 8/86 cases (DD-78, 111, DJA-156, 270, 329\footnote{\label{note_dja329}DJA-329 is characterized by a particularly complex kinematic structure in the 1D spectrum \citep{d'eugenio_2024}. We modeled the gas emission as a combination of a narrow component for all emission lines that shared the velocity offset with a broad \ha\ component, plus an outflowing component for \oiii\ and \nii. The broad component models well the extended red wing of the \ha+\nii\ complex, but it is sensitive to the initial guesses. The resulting \mbh\ could be overestimated by 1 dex and should be thus used with caution. Also, the significance of the \nev\ detection floats in the $2.8-3.1\sigma$ range depending on the exact kinematic modeling. We consider it a robust detection in the remainder of this work. A similar case is DJA-700, not detected in \nev.}, 616, 700, 859), a narrow gas emission component was insufficient to reproduce the wide wings in the observed spectra. We thus added a broad component for \ha, \hb, and \hg\ lines with a velocity centroid tied to that of the narrow component. In six cases, we measured $\rm{FWHM}\gtrsim4000$ \kms, indicative of accreting SMBHs.\footnote{The broad \ha\ detections are not an artifact of the spectral S/N around the line: they are found across a wide range of continuum S/N near \ha\ (S/N $\sim 10-25$), with no preference for the highest-quality spectra. The same holds for the \nev\ detections, which span S/N $\sim 2-8$ around the line.} In the remaining two spectra, we detect faint extended wings with $\mathrm{FWHM}\approx 2000 $ \kms\ at $>3\sigma$ significance, which may indicate actively accreting intermediate-mass black holes ($M_{\rm BH}\approx10^7\,M_\odot$, see below) given their stellar mass -- but are also compatible with typical outflow velocities driven by nuclear activity. In the remainder of the analysis, we flag these sources as possibly hosting accreting intermediate-mass black holes, though this classification remains uncertain. 
A detailed analysis of black hole mass measurements and scaling relations for the full DeepDive and archival samples is deferred to a dedicated work (Y. Shibanuma et al., in prep.). For the rest of the analysis, we add to our pool of galaxies the X-ray detected, quiescent source at $z=2.094$ analyzed in detail in \cite{ito_2025_agn}, where a broad line component with $\mathrm{FWHM}=4365\pm81$ \kms\ is necessary to model the \ha\ and \hb\ emission.

Our total sample of QGs with \nev\ coverage thus comprises 87 systems, out of which seven have an empirical estimate of \mbh\ from broad \ha\ lines ($\rm{FWHM}\gtrsim4000$ \kms) and two possibly host intermediate-mass black holes ($\mathrm{FWHM}\approx2000$ \kms).

\section{Results}
We detect \nev\ emission in 6 (10) sources at $>3\sigma$ ($>2\,\sigma$) (Table \ref{tab:table}). We show the spectra of the 5 newly identified sources with \nev\ detections\footnote{See footnote \ref{note_dja329}.} from the literature compilation presented in \cite{ito_2025_deepdive} in Figure \ref{fig:spectra}, while the sixth galaxy is shown in Figure 2 of \cite{ito_2025_agn}. This source, together with DD-78, DD-111, and DJA-329, shows typical colors and morphologies of massive quenched systems at high redshift. DD-82 displays some properties reminiscent of a ``little red dot'' (LRD, \citealt{matthee_2024}): it is compact, yet resolved when modeled with a S\'ersic profile ($R_{\rm eff} = 0.41^{+0.06}_{-0.05}$ kpc, $n=1.3^{+0.4}_{-0.3}$ at $\sim4000$ \AA\ rest-frame); it shows a shallow Balmer break and red colors, but does not fall within common photometric color selections \citep{kokorev_2024_lrds}; it displays bright emission lines, but broad features under \ha\ and \hb\ are only marginally detected, while residuals around the narrow \oiii\ doublet may suggest a broad component associated with an outflow; the continuum shows weak absorption features, notably the G band+\hg\ complex, of likely stellar origin; finally, the strong \nev\ emission itself is not a common trait of LRDs. We retain this galaxy in our sample, with the caveat that it may be a hybrid object in which host and AGN emission remain significant while the LRD component is fading (see \citealt{kokorev_2024} for another possible example). DJA-924, on the other hand, shows a radically different morphology: the spectrum traces the core of a bulge-dominated spiral, yet the total photometry, including the star-forming disk, returns an SFR well below the main sequence at $z=2$, with rest-frame $UVJ$ colors consistent with classical quiescent galaxy selections \citep{ito_2025_deepdive}. We retain this object in our analysis, though the mechanism triggering the nuclear activity may differ from the rest of the sample. We note that 84/87 galaxies have spectral coverage of the \oiii\ emission line, which is detected at $>3\sigma$ ($>2\sigma$) in 32 (37) sources, including all \nev-detected ones. We return to the \oiii\ -- \nev\ relation, \oiii\ being a classical tracer of AGN bolometric luminosity \citep{heckman_2004}, in Section \ref{sec:scaling_relations}.

% Figure: BPT diagram
\begin{figure}
    \centering
    \includegraphics[width=\columnwidth]{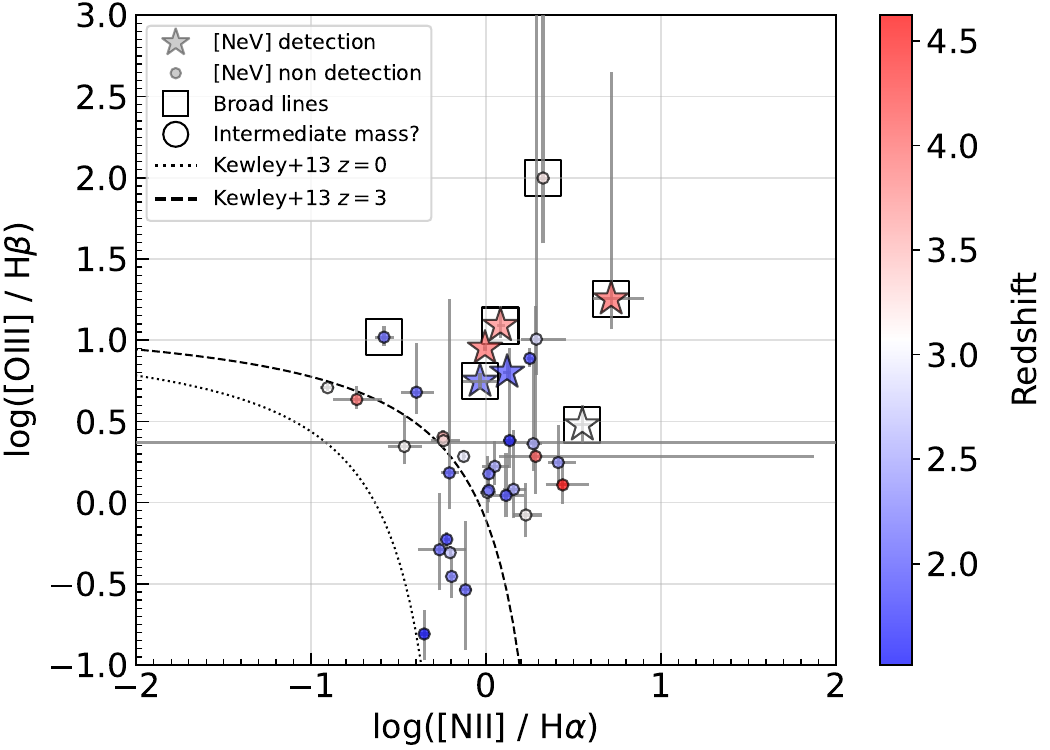}
    \caption{Distribution of QGs with \nev\ coverage in the BPT diagram. Stars indicate \nev\ detections, while circles mark upper limits. The symbols are color-coded according to redshift. Open black squares and circles indicate the sources with broad and intermediate lines, respectively. The dotted and dashed black lines mark the SF-AGN separation lines at $z=0$ and $z=3$, respectively, following \cite{kewley_2013p}.}
    \label{fig:bpt}
\end{figure}
% Figure: [NeV] vs [OIII] emission
\begin{figure}
    \centering
    \includegraphics[width=\columnwidth]{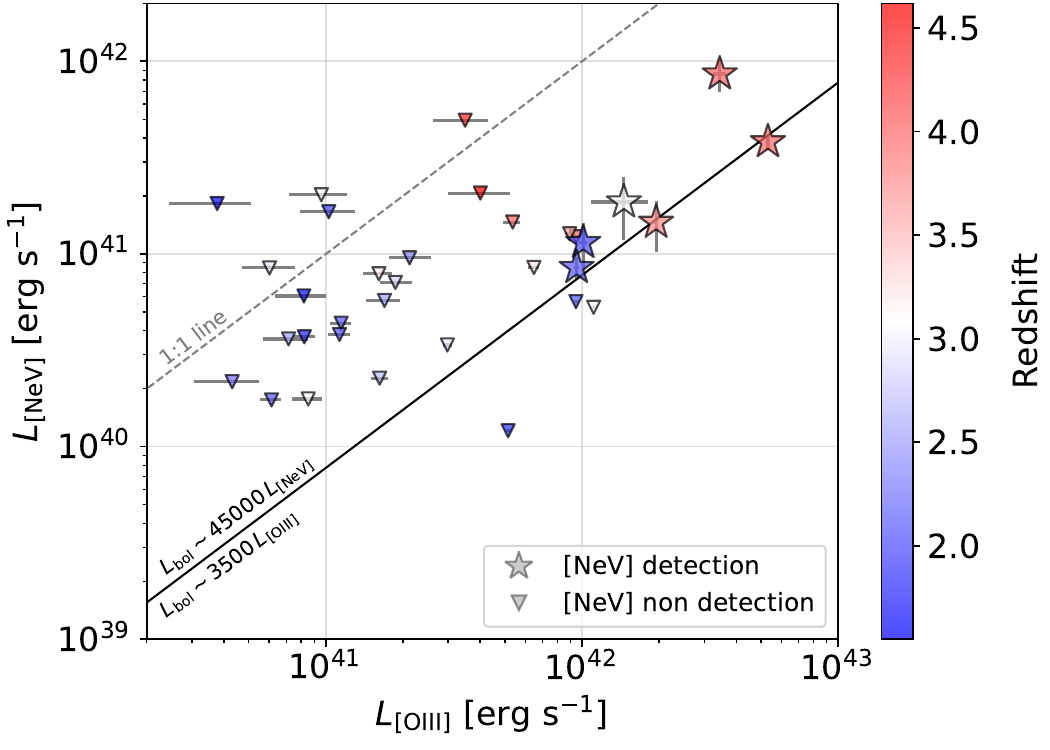}
    \caption{Relation between observed \nev\ and \oiii\ luminosities. Filled stars indicate detections of both lines in our sample, while triangles indicate upper limits on \lnev. The symbols are color-coded according to redshift. The dashed and solid lines indicate the one-to-one relation and the expected ratio based on the bolometric corrections adopted in this work \citep{heckman_2004,reiss_2025}.}
    \label{fig:nev_oiii}
\end{figure}
\subsection{\nev\ traces nuclear activity in quenched galaxies}
We show the redshift and \nev\ luminosity ($L_{\rm [NeV]}$) distributions of our galaxies, along with samples of \nev-detected sources from the local universe to $z=7$ \citep{gilli_2010, cales_2013, cleri_2023, chisholm_2024, scholtz_2025, trakhtenbrot_2025, barchiesi_2025} in Figure \ref{fig:z_nev}. The effect of the inhomogeneous coverage and Malmquist bias is reflected in the distribution of the upper limits as a function of redshift. We detect \nev\ at $>3\sigma$ in 4 out of 7 unambiguous broad-line AGN, while the remaining detections are in narrow-line systems. All the brightest X-ray detections with rest-frame $L_{\rm 2-10\,keV}\gtrsim10^{44}$ \es\ are also detected in \nev\ with $L_{\rm 2-10\,keV}/L_{\rm [NeV]}$ ratios of $\approx300-1000$. Notably, these are all broad-line AGN, complementing the X-ray-obscured Type 2 AGN in recent works \citep{barchiesi_2024, barchiesi_2025}. The rest of the narrow-line sample with \nev\ detections remain undetected in X-ray. The $3\sigma$ upper limits on the $L_{\rm 2-10\,keV}/L_{\rm [NeV]}$ ratios of $<80-200$ are consistent with the obscured AGN population identified in previous works \citep{gilli_2010, barchiesi_2024}. Figure \ref{fig:bpt} shows the location of the \nev\ sample in the classical BPT diagram \citep{baldwin_1981}. All sources exhibit high \nii/\ha\ and \oiii/\hb\ ratios indicative of AGN activity when compared with redshift-dependent classical selection criteria \citep{kewley_2013p}. The \nev\ detections show some of the most extreme line ratios in the overall population. \nev\ is thus a reliable tracer of the most extreme, possibly obscured, AGN activity in distant quenched galaxies.

\subsection{Scaling relations and quasar-like activity}
\label{sec:scaling_relations}
% Figure: Black hole scaling relations
\begin{figure*}
    \centering
    \includegraphics[width=\textwidth]{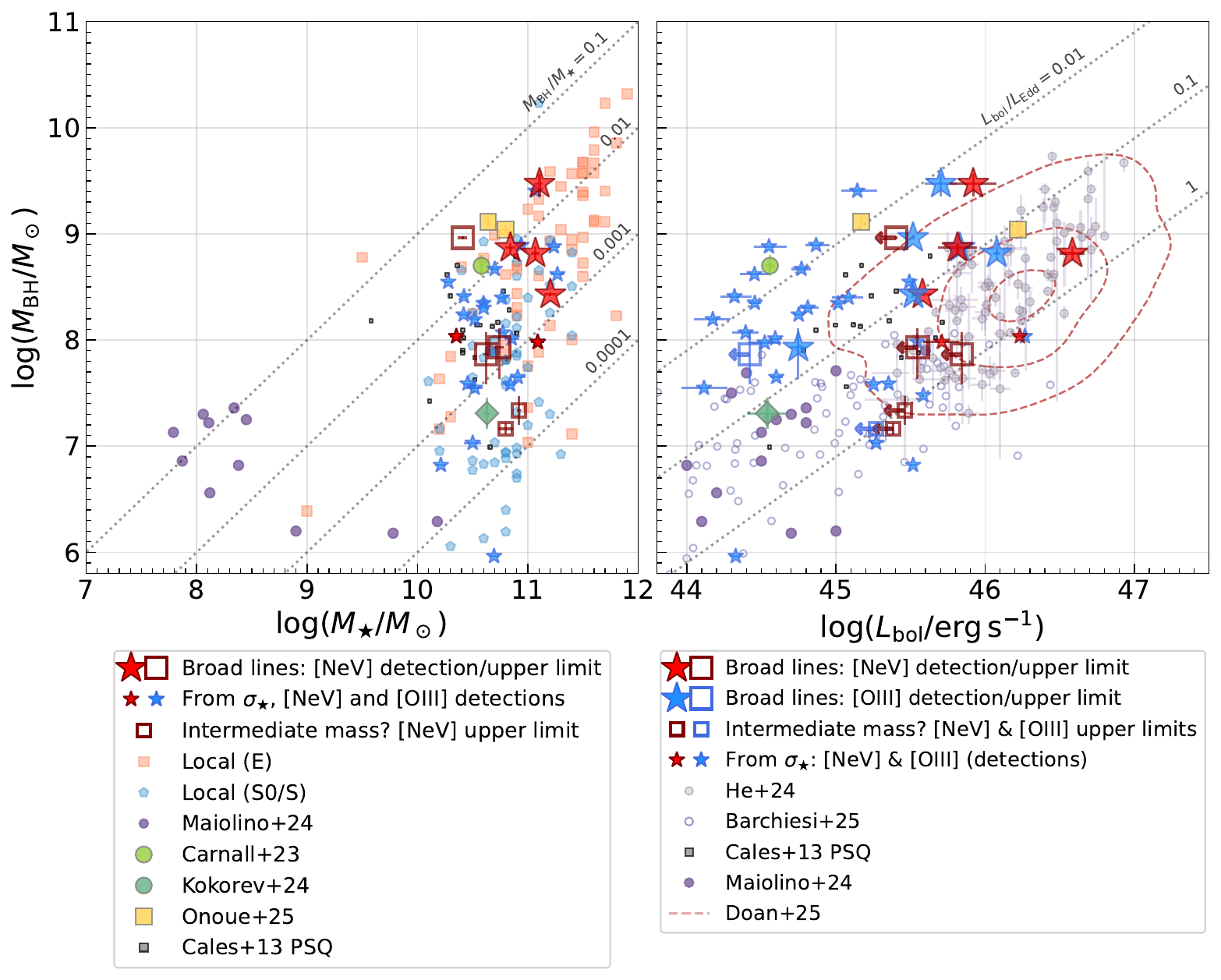}
    \caption{
    \textit{Left:} \mstar--\mbh\ relation. Large red symbols indicate our sample of QGs with broad ($\mathrm{FWHM}\gtrsim 4000$ \kms) \ha\ emission, in which filled stars and open squares show the \nev\ detections and upper limits, respectively. 
    Small empty squares mark our QGs with intermediate ($\mathrm{FWHM} \approx 2000$ \kms) \ha\ emission, undetected in \nev. 
    Small filled red and blue stars show the predicted location in the \mstar--\mbh\ plane of \nev- and \oiii-detected QGs without broad line region signatures, assuming the \mbh--\sigmastar\ relation from \cite{mcconnell_2011}. 
    Local ellipticals (E) and S0/spiral galaxies (S0/S) are shown as light orange squares and blue pentagons, respectively \citep{kormendy_2013, greene_2020}. Post-starburst QSOs at $z=0.3$ from \cite{cales_2013} are shown as small gray squares. 
    Overmassive black holes in low-mass galaxies at $4<z<7$ from \cite{maiolino_2024} are shown as filled violet circles. 
    The quenched systems at $z=4.6$ and $z=4.1$ from \cite{carnall_2023b} and \cite{kokorev_2024} are shown as a green circle and a diamond, respectively. 
    The QSO-selected, post-starburst systems in \cite{onoue_2025} are shown as with yellow squares. 
    The dotted lines indicate constant \mbh/\mstar\ ratios from 0.1 to $10^{-4}$. 
    \textit{Right:} \lbol--\mbh\ relation. The meaning of large and small red and blue symbols is as in the left panel and in the legend. The light purple and dark empty circles show the low-luminosity QSO sample at $z=4$ from \cite{he_2024} and the \nev-detected low-redshift AGN sample from \cite{barchiesi_2025}. Low-mass galaxies at $4<z<7$ from \cite{maiolino_2024} are shown as filled violet circles and the PSQ in \cite{cales_2013} as small gray squares. The red dashed contours enclose the 10th, 50th, and 90th percentiles of the \nev-detected Type 1 AGN from SDSS in \cite{doan_2025}. The dotted lines indicate constant Eddington ratios (\lbol/$L_{\rm Edd}$) of 1, 10, and 100\%.}
    \label{fig:bh_relations}
\end{figure*}
For the sources with detected \ha\ broad line emission, we derive the black hole mass as in \cite{reines_2013}:
\begin{equation}
    \begin{split}
    \mathrm{log}(M_{\rm BH}/M_\odot) = 6.57 + \mathrm{log}(\epsilon) + 0.47\,\mathrm{log}(L_{\rm H\alpha,\,broad}/10^{42}\, &\mathrm{erg\,s^{-1}})\\
    +2.06\,\mathrm{log}(\mathrm{FWHM_{H\alpha,\,broad}/1000\,km\,s^{-1}})
    \end{split}
\end{equation}
where $L_{\rm H\alpha,\,broad}$ and $\mathrm{FWHM_{H\alpha,\,broad}}$ are the luminosity and width of the broad \ha\ component, and $\epsilon=1.075$ is a geometric correction factor \citep{reines_2015}. For sources without a significant broad line component, we assumed the \mbh--\sigmastar\ relation in \cite{mcconnell_2011}. We derived the bolometric luminosities from the \nev\ emission as in \cite{reiss_2025} ($\mathrm{log}(L_{\rm bol}/\mathrm{erg\,s^{-1}}) = \mathrm{log}(L_{\rm [NeV]}/\mathrm{erg\,s^{-1}})+4.65$) and from the \oiii\ line following the calibration in \cite{heckman_2004} ($\mathrm{log}(L_{\rm bol}/\mathrm{erg\,s^{-1}}) = \mathrm{log}(L_{\rm [OIII]}/\mathrm{erg\,s^{-1}})+3.54$), finding broad agreement whenever both \nev\ and \oiii\ are robustly detected, with the observed $L_{\rm [OIII]}\sim 12 \times$ brighter than $L_{\rm [NeV]}$ (Figure \ref{fig:nev_oiii}). We then computed the Eddington luminosity as $L_{\rm Edd} / \mathrm{erg\,s^{-1}} = 1.3\times10^{38}(M_{\rm BH}/M_\odot)$, and the Eddington ratio $\lambda_{\rm Edd} = L_{\rm bol}/L_{\rm Edd}$ as a proxy of the radiative efficiency of the central black holes. We caution, however, that these relations are calibrated locally and their applicability at high redshift remains uncertain (e.g., see \citealt{abuter_2024, newman_2025} for new dynamical estimates of \mbh\ in distant galaxies).

Figure \ref{fig:bh_relations} shows these properties in the context of the well-known scaling relations among \mbh, \mstar, and \lbol. For context, we include reference samples across redshift where available. The literature galaxies include local ellipticals, S0, and spirals \citep{kormendy_2013, greene_2020}, high-redshift black holes in low-mass systems \citep{maiolino_2024}, low-luminosity QSOs at $z=4$ \citep{he_2024}, and galaxies more akin to our sample of quenched galaxies \citep{carnall_2023b, kokorev_2024, onoue_2025}. 

The \mbh\ measurements place our QGs on the known scaling relation with \mstar\ observed in the local universe, albeit within its large scatter \citep{kormendy_2013, greene_2020}, in contrast to the overmassive black holes found in bluer galaxies at lower stellar masses, and in agreement with what was previously reported in \cite{ito_2025_agn}. These systems are also broadly consistent with the \mbh--\sigmastar\ relation (Y. Shibanuma et al. in prep.), which supports our assumption of using the local relation from \cite{mcconnell_2011} to estimate \mbh\ for sources without direct measurements from broad lines.
Moreover, the bolometric luminosities derived from \nev\ and \oiii\ indicate that approximately half of the \nev-detected sample has high Eddington ratios ($\lambda_{\rm Edd} = L_{\rm bol}/L_{\rm Edd}=10-50$\%, Figure \ref{fig:bh_relations}). These values are consistent with those observed in the low-luminosity tail of QSOs at $z=4$ \citep{he_2024}, and they are higher than the values reported by \cite{onoue_2025} for a post-starburst system at $z=6.5$. These values are also similar to those of \nev-detected Type 1 AGN in the local universe \citep{doan_2025} and at Type 2 AGN at cosmic noon \citep{barchiesi_2025}. The rest of the sample detected in \oiii, but not in \nev, typically shows Eddington ratios of a few percent. This confirms that \nev\ detections cleanly select the brightest and most powerful tail of radiatively efficient AGN in our sample of distant quenched galaxies.

\subsection{Black hole growth after quenching}
\begin{figure}
    \centering
    \includegraphics[width=\columnwidth]{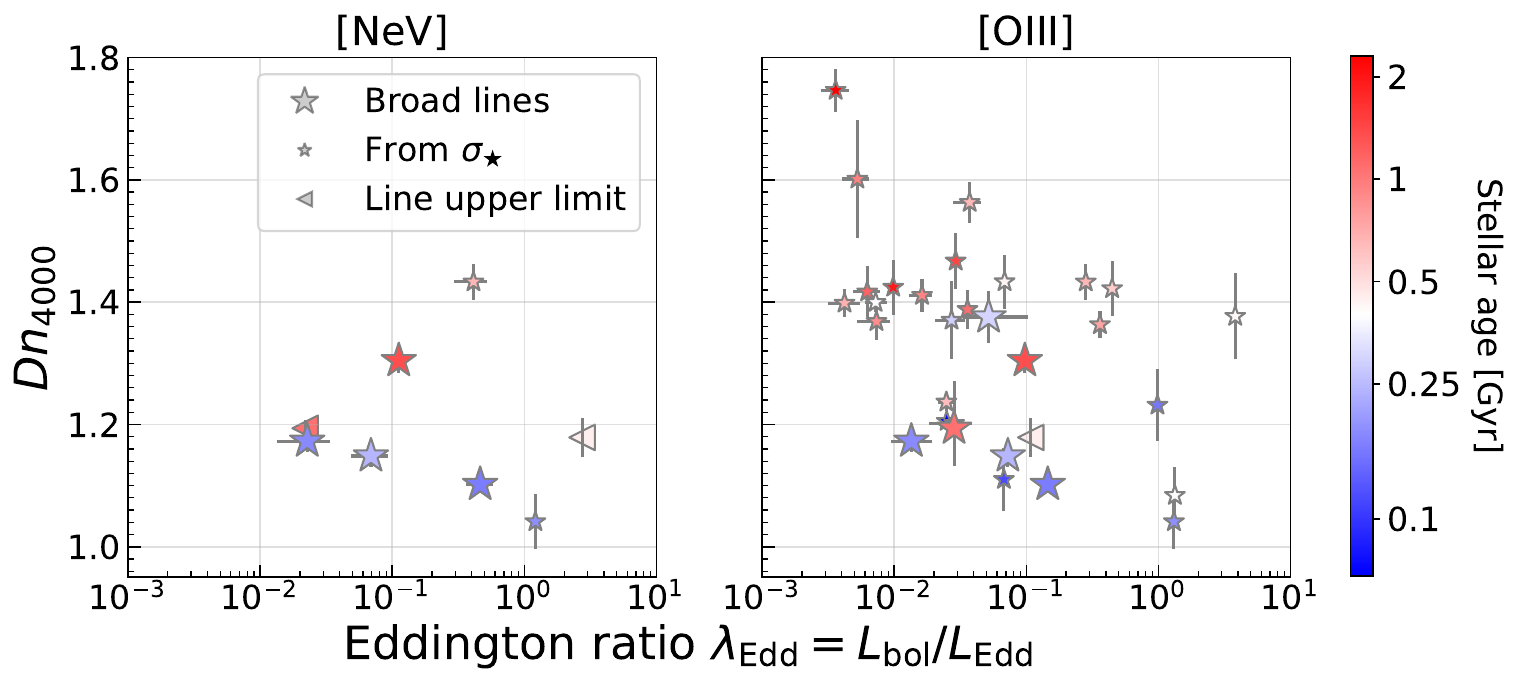}
    \caption{
    \dn\ as a function of the Eddington ratio \eddratio. The left and right panels show \lbol\ derived from \lnev\ and \loiii, respectively. Large symbols indicate galaxies with \mbh\ measurements from broad lines and, thus, direct estimates of $\lambda_{\rm Edd}$. Small symbols mark the expected location of QGs assuming they lie on the local $M_{\rm BH}--\sigma$ relation from \cite{mcconnell_2011}. The symbols are color-coded according to the light-weighted stellar age from the modeling with pPXF.}
    \label{fig:eddratio_dn}
\end{figure}

Our findings also imply that sources a few hundred Myr after their main formation and quenching epoch can still harbor highly accreting and luminous central black holes. Single-epoch data are insufficient to establish whether we are witnessing the fading of radiatively efficient AGN activity and a transition to low-efficiency radio maintenance modes predicted by standard evolutionary models \citep{hopkins_2006} or a recent flare-up or decay of QSO-like activity triggered by episodic accretion of gas (especially in the case of DD-924, given its morphology) or mergers. The strongest \nev\ emitters are found at the low-\dn\ end of the distribution ($D_{\rm n}4000\approx1.2$), suggesting a prevalence in post-starburst, rather than old quenched galaxies. This is broadly consistent with the detection in ``green valley'' galaxies reported in previous works at lower redshift \citep{vergani_2018, vergani_2025, barchiesi_2025}. Although based on only a handful of empirical \mbh\ measurements from broad lines, Figure \ref{fig:eddratio_dn} shows that the highest $\lambda_{\rm Edd}$ values are found at the lowest \dn, corresponding to younger ages and more recent quenching. The remaining points increase the statistical sample but also introduce scatter in the possible correlation, given that their black hole masses, and thus Eddington luminosities, are inferred from the scattered \mbh--\sigmastar\ relation. We also note that a redshift selection bias, whereby only the brightest broad emission lines (and thus the highest \mbh\ and $L_{\rm Edd}$) and the youngest ages (lowest \dn) are detectable in the most distant systems, would introduce a spurious correlation perpendicular to the tentative trend in Figure \ref{fig:eddratio_dn}.

In absolute terms, a small mass of gas might be sufficient to trigger intense black hole accretion, while leaving the total stellar mass almost completely unaffected, given the typical mass of the targets in our sample (Figure \ref{fig:m_dn}). We quantified the black hole accretion rate (BHAR) in our galaxies as follows:

\begin{equation}
    \mathrm{BHAR}\,[M_\odot\, \mathrm{yr}^{-1}] = L_{\rm bol}  \frac{1-\epsilon}{\epsilon c^2}
\end{equation}

where $\epsilon=0.1$ is the assumed mass-to-radiation conversion efficiency and $c$ the speed of light. We computed this quantity from $L_{\rm bol}$ derived from \nev\ and \oiii. We then compared the BHAR with the SFR from the SED modeling, separating the sample into \nev-detected and \oiii-detected subsets. In the most extreme cases, such as DD-78, we find $\mathrm{BHAR/SFR}>1$, suggesting that the central black holes undergo cycles of growth spurts. Nevertheless, the absolute BHAR is of the order of a few \myr---a quantity that, if converted into stars, would result in negligible stellar mass growth.

As among the most energetic and extreme phenomena observed, such growth is unlikely to be sustained. We derived an order-of-magnitude duty cycle for the AGN activity by computing the fraction of \nev-detected sources in the full sample (6/87) spanning the $z=1.52-4.66$ range: $t_{\rm duty} = N_{\rm det}/N_{\rm tot} \times \Delta t(z=1.52-4.66) = 199^{+109}_{-53}$ Myr. Uncertainties in the detection fractions and the corresponding duty cycles are given by binomial statistics with two-sided 68\% confidence intervals following \cite{cameron_2011}.
Including only the brightest sources with $L_{\rm bol}>10^{46}$ \es, this timescale decreases by approximately a factor of two ($t_{\rm duty} = 66^{+83}_{-21}$ Myr based on a detection fraction of 2/87, similar to the Salpeter timescale for AGN accretion). We note that only a small fraction of AGN in the local Universe emit detectable \nev\ (e.g., $\sim5$\% of Type1 AGN in SDSS; \citealt{doan_2025}), possibly due to dust obscuration \citep{mckaig_2024}. By construction, our quiescent sample has low galaxy-integrated dust extinction, which may facilitate the detection of \nev. 
 Our estimates for the duty cycle of the AGN activity are consistent with the values for QSOs at similar redshifts \citep[and references there-in]{arita_2025}. They are also consistent with the loose upper limit that we can derive from the location of our targets on the \mbh--\mstar\ relation. DD-78, the prime example of our \nev\ detections and quasar-like activity in a post-starburst galaxy at $z=4.01$, could serve for a back-of-the-envelope calculation. Under the simple assumption that DD-78 keeps growing at the current $\rm  BHAR_{[NeV]} = 6.1 \pm 1.2$ \myr\ and $\rm SFR_{SED} < 23$ \myr\ (at $3\sigma$), and considering that the source is already on the local \mstar--\mbh\ relation, it would take $<525$ Myr to reach a 3\% \mbh/\mstar\ ratio, marking the upper limit of the relation.
% Figure: Eddington ratio vs Dn4000

\section{Discussion}
% Figure: Mstar, MBH, Eddington ratio distributions for models and simulations
\begin{figure*}
    \centering
    \includegraphics[width=\textwidth]{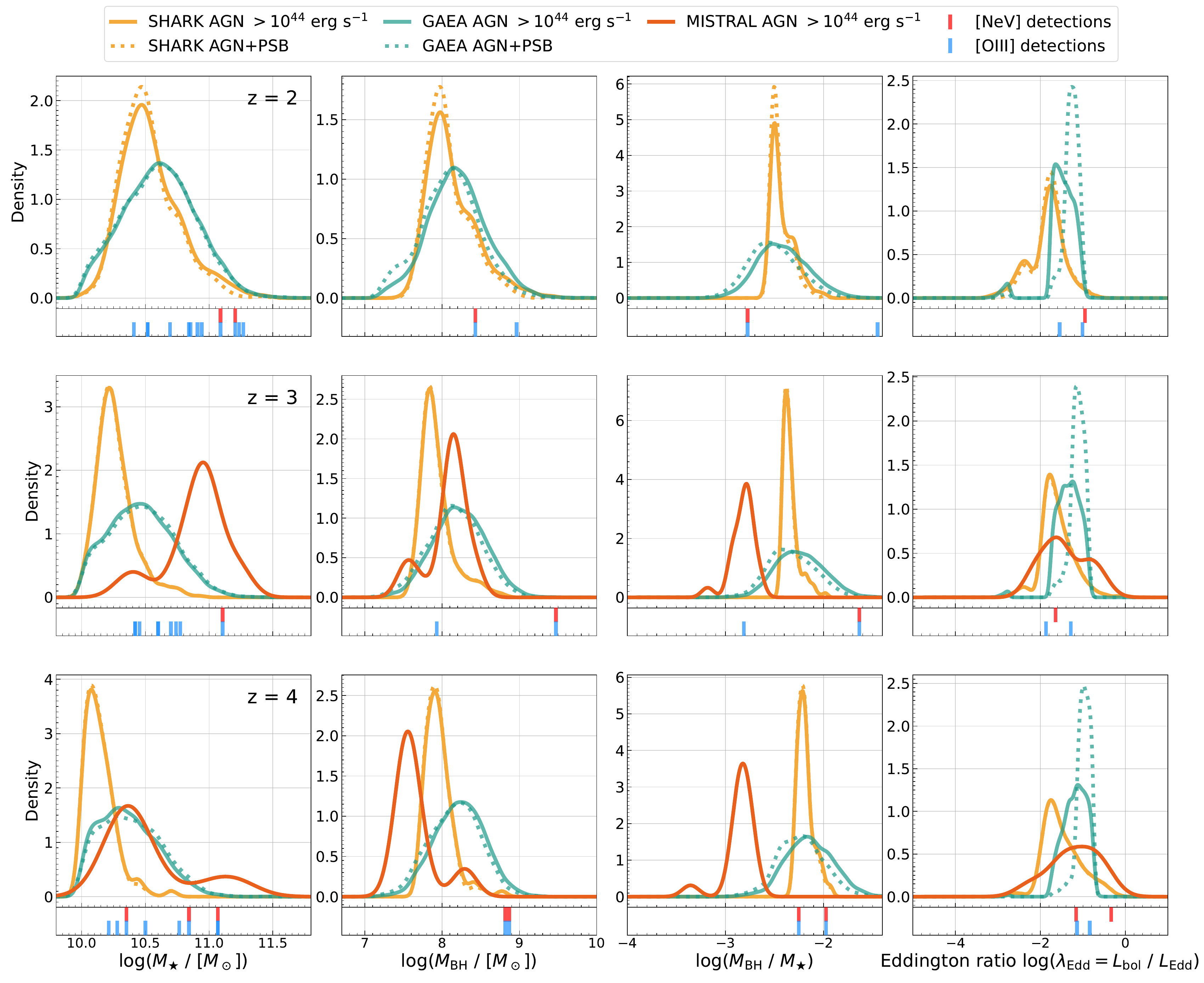}
    \caption{Stellar masses, black hole masses, their ratios, and Eddington ratios for the \shark\ (solid yellow lines), \gaea\ (solid teal), and {\sc Black Dawn}/\mistral\ (solid red) quiescent galaxies ($\mathrm{sSFR} < 0.2/t_{\rm Hubble}$) with AGN activity ($L_{\rm bol}>10^{44}$ \es) at $z=2-4$, as labeled. For \shark\ and \gaea, dotted lines show the recently quenched subsample ($t_{\rm quench}<250\,\mathrm{Myr}$). For {\sc Black Dawn}/\mistral, each post-quenching AGN burst is treated as an individual measurement. 
    Under each panel, we mark the observed properties of our \nev- and \oiii-detected samples in red and blue, respectively. The number of sources with an empirical \mbh\ estimate from broad lines, and thus $L_{\rm Edd}$, is much lower than those with an \mstar\ estimate, resulting in sparse observational data points in the three rightmost panels.
}
    \label{fig:model_distributions}
\end{figure*}
\subsection{Quasar-like emission in post-starburst galaxies}
The detection of \nev\ in massive quenched systems at $z\sim1.5-4.5$ confirms that quasar-like activity can persist a few hundred Myr after quenching. Figure \ref{fig:spectra} shows that the spectra of our sources are dominated by the stellar continuum. This is by construction: our selection is inherently galaxy-based rather than AGN/QSO-based, built on strong Balmer breaks, $UVJ$ colors, and low sSFR.

The coexistence of post-starburst stellar features in bright AGN and, vice versa, of AGN emission in recently quenched galaxies has been reported in the local universe, where ground-based spectra provide signal-to-noise ratios sufficient to disentangle the two components. A class of local ``post-starburst QSOs'' (PSQs), selected on the basis of broad line emission and deep \hd\ absorption, has been the subject of study over the past three decades \citep{brotherton_1999, cales_2013, cales_2015}. We show the location of a sample of $33$ PSQ with \nev\ detections at $z=0.3$ from \cite{cales_2013} as dark gray squares in Figure \ref{fig:bh_relations}. For this comparison, we re-derived the bolometric luminosity from \nev\ and \mbh\ using the same prescriptions as for our sample to reduce systematics. These galaxies occupy a broadly similar region of parameter space as our \nev-detections, with the lower average \mstar, \mbh, and \lbol\ likely reflecting a combination of Malmquist bias and the scarcity of post-starburst galaxies at low redshift \citep{wild_2016}. The main difference is that we do not include bluer QSOs and do not require Type 1 broad lines. Interestingly, the brightest AGN emission in local PSQs tends to be associated with early-type galaxies and recent major mergers, consistent with the standard scenario in which a galaxy collision triggers a starburst and nuclear activity \citep{ellison_2019_agn_mergers, ellison_2022_mergers_quenching, ellison_2025_agn_mergers}. Fainter QSOs, by contrast, reside in spirals and are more likely driven by secular mechanisms \citep{cales_2013}. 

Among our \nev\ detections, the brightest source, DD-78, shows clear broad emission and a spheroidal morphology with a disky S\'{e}rsic index \citep{tanaka_2019, ito_2024, kakimoto_2026_deepdive} and a possible velocity gradient in the stellar kinematics, albeit uncertain given the compact size and slit alignment (Ito et al. in prep.). The galaxy is isolated, with no close photometric companions or morphological disturbances. DJA-329 also shows strong rotational support and is surrounded by low-mass gaseous satellites, which may help fuel the central SMBH and trigger outflows \citep{d'eugenio_2024}. Both objects may thus partially deviate from the major merger formation channel commonly associated with local post-starburst QSO hosts, which typically leave behind compact, pressure-supported stellar structures. However, the fact that the kinematic and structural properties typical of rotationally supported systems were produced by past major mergers cannot be ruled out -- and mergers (minor or major) are among the mechanisms triggering quenching in the distant universe, where the merger rates are higher \citep[e.g.,][]{xie_2024}. Moreover, in models such as \gaea\ (Section \ref{sec:models}), secular processes are a viable channel for triggering QSO activity and producing bright AGN in quenched disks. 

Compared with the faint end of QSOs at even higher redshifts, the galaxies presented here may be the natural evolved counterparts of the PSQ-like systems at $z>6$ in \citet{onoue_2025} and the photometric candidates in \citet{ding_2025_shellqs}. Moreover, some sources may also be traversing intermediate phases in which the central SMBH is enshrouded in the dense gas cocoons proposed to explain LRDs \citep{kokorev_2024, rusakov_2026}, though this connection remains speculative given the poorly understood nature of LRDs and their host galaxies. DD-82 may represent such a case. For DJA-924, its spiral morphology---though dominated by a central bulge and characterized by a low overall SFR---points toward turbulence, bar-driven inflows, or other secular processes as the likely drivers of its ongoing low-luminosity AGN activity, rather than a major merger.

The degree of overlap between QSO emission and evolved stellar populations remains debated and depends on the adopted selection criteria and AGN proxies. Results also differ depending on the approach: spectroscopy allows a cleaner AGN--host separation and more precise dating of stellar populations but is typically limited to small samples, while large photometric surveys offer statistical power at the cost of greater uncertainties in stellar population modeling. The relation is not symmetric: as noted above, an AGN-based approach does not necessarily sample the same population as a galaxy-based one. In the local universe, a disproportionately high fraction of QSOs relative to mass-controlled samples are hosted in rare photometrically selected post-starburst galaxies, predominantly associated with gas-rich major mergers \citep{krishna_2025}. How common QSO activity is within the post-starburst population up to cosmic noon remains an open question \citep{yesuf_2014, pawlik_2018, almaini_2025}, complicated by the relative timing of the starburst and AGN onset \citep{Wild_2010, ellison_2025_agn_mergers} and by the degree of dust obscuration of the central AGN. Consistent with this picture, stacking of optical spectra at $z\sim0.6-1.2$ reveals \nev\ emission preferentially in galaxies with blue colors but a recent shutdown of star formation, pointing at the existence of young, active post-starburst systems \citep{vergani_2018}. Starting from an X-ray selection of obscured Type 2 AGN, \nev\ detections have also been associated with galaxies at the end of their star-forming phase or shortly after its cessation, based on photometric SEDs \citep{barchiesi_2024, barchiesi_2025}. Our approach is complementary: we selected bona fide quenched galaxies and subsequently searched for \nev\ emission and AGN activity. Crucially, access to the full rest-frame optical spectrum for each object allows us to model the ionized gas emission and stellar populations on an individual galaxy basis, without relying on stacking, at the cost of small-number statistics. Larger, more complete spectroscopic samples of recently quenched galaxies at high redshift will be essential to make progress on this front. Signatures of AGN activity, traced by ionized gas emission \citep{bugiani_2025, baker_2025, stevenson_2026} and feedback \citep{davies_2024, belli_2024, d'eugenio_2024, wu_2025, valentino_2025, taylor_2026, zhu_2026_outflows}, appear nearly ubiquitous in such systems. In this context, targeting high-ionization lines such as \nev\ offers a particularly clean diagnostic: it isolates the brightest tail of AGN emission free from contamination by residual star formation, and remains well within reach of \textit{JWST}/NIRSpec in wavelength coverage and sensitivity, as this study demonstrates.

\subsection{Models of AGN feedback and quenching in the distant Universe}
\label{sec:models}
 
Finally, we compare predictions from theoretical models of galaxy formation, specifically \mstar, \mbh, and \lbol, with our observational findings, with the aim of constraining their different subgrid feedback implementations. Although our sample is not complete given its archival nature, it spans a representative parameter space for quenched galaxies at $z>1.5$ and provides a useful benchmark for blind \nev\ detections in such systems.
 
In particular, in our work we compare the semi-analytical \shark\ model \citep[v2.0,][]{lagos_2024} and the GAlaxy Evolution and Assembly model \citep[\gaea, in the version presented in][]{delucia_2024} with a set of hydrodynamical zoom-in simulations evolved using the fiducial ``stochastic'' version of the \mistral\ AGN feedback model \citep{farcy_2025}. The latter are selected among the 50 most massive galaxies of the Illustris \tng-100 simulation at $z=3$, and are resimulated at twice the spatial resolution and with improved snapshot sampling, forming part of the {\sc Black Dawn} sample (Farcy et al., in prep.). The large statistics afforded by \shark\ and \gaea\ capture population trends at each snapshot, while the high spatial resolution of the \blackdawn\ zoom-in simulations allows for a detailed follow-up of the variation of stellar and black hole mass growth. 

The AGN feedback implementations in \shark\ and \gaea\ differ substantially in both their physical assumptions and the way they connect black hole growth to galaxy quenching. \shark\ (dark matter particle mass $M^{\rm part}_{\rm DM}=2.21 \times 10^8 h^{-1}$ \msun) adopts a two-mode AGN feedback model consisting of a jet (``radio'') mode and a radiatively driven wind (``QSO'') mode. The jet mode suppresses cooling flows in halos that have developed a hot atmosphere, with the jet power depending explicitly on the black hole mass, spin, and accretion rate. In the QSO mode, \shark\ includes radiation-pressure-driven winds that can expel gas from galaxies and even from their halos if the outflow energy exceeds the halo binding energy. However, the latter is not responsible for the bulk of the massive quenched galaxies in the model, which are instead primarily quenched through the jet mode. 
For \gaea, we consider a realization run on the Millennium Simulation (\citealt{delucia_2024}, corresponding to a dark matter particle resolution of $M^{\rm part}_{\rm DM}=8.6 \times 10^8  h^{-1}$ \msun). This model assumes that AGN are triggered by disk instabilities and/or mergers, which cause a fraction of the cold gas in the model galaxies to lose angular momentum and fall toward the center, where it can be accreted onto the central black hole on a viscous accretion timescale. Accretion powers QSO-driven outflows, whose mass loading factor scales empirically with the AGN bolometric luminosity. Therefore, despite \gaea\ also including a radio-mode feedback prescription that is mainly effective in massive haloes at lower redshift, the quenching of massive galaxies in the early universe is associated with the QSO mode. This is phenomenologically similar to the outcome of the \mistral\ feedback model, in which the quenching of massive galaxies at $z > 3$ is driven by QSO-driven winds. \mistral\ adopts a stochastic kinetic AGN feedback model in which accreting black holes launch bipolar winds aligned with the angular momentum of the surrounding gas. The model couples AGN energy to the black hole's immediate surroundings through stochastic momentum injection events that drive large-scale outflows that suppress gas accretion, rather than heating quasi-static hot halos. The accretion rate in \blackdawn\ ($M^{\rm part}_{\rm DM}=5.3 \times 10^5 h^{-1}$ \msun) uses an Eddington-limited Bondi accretion scheme, and \mistral\ is used at all accretion regimes.

For all simulations, we selected quiescent galaxies with ongoing AGN activity using a threshold of $\rm sSFR \leq 0.2/t_{\rm Hubble}$ and $L_{\rm bol}>10^{44}$ \es (the lower end of the values inferred from \nev\ and, for reference, \oiii\ in our observed sample). Imposing a bolometric luminosity threshold skews the distribution toward the most recently quenched objects in each snapshot, as discussed below. Indeed, the vast majority of quiescent galaxies with this level of AGN activity quenched less than $t_{\rm quench}<250$ Myr before the snapshot at which they are selected.\footnote{Here we define $t_{\rm quench}$ as the time since a galaxy last crossed the sSFR threshold to become quiescent (\shark, \blackdawn), or since the peak of the AGN activity that drove its quenching (\gaea). While both definitions broadly capture the same physical process, small offsets may arise between them and with uncertain observational estimates derived from SFH modelling. For the purpose of identifying broad trends with $t_{\rm quench}$ in models, this does not significantly affect our conclusions.}
 \begin{figure*}
    \includegraphics[width=\textwidth]{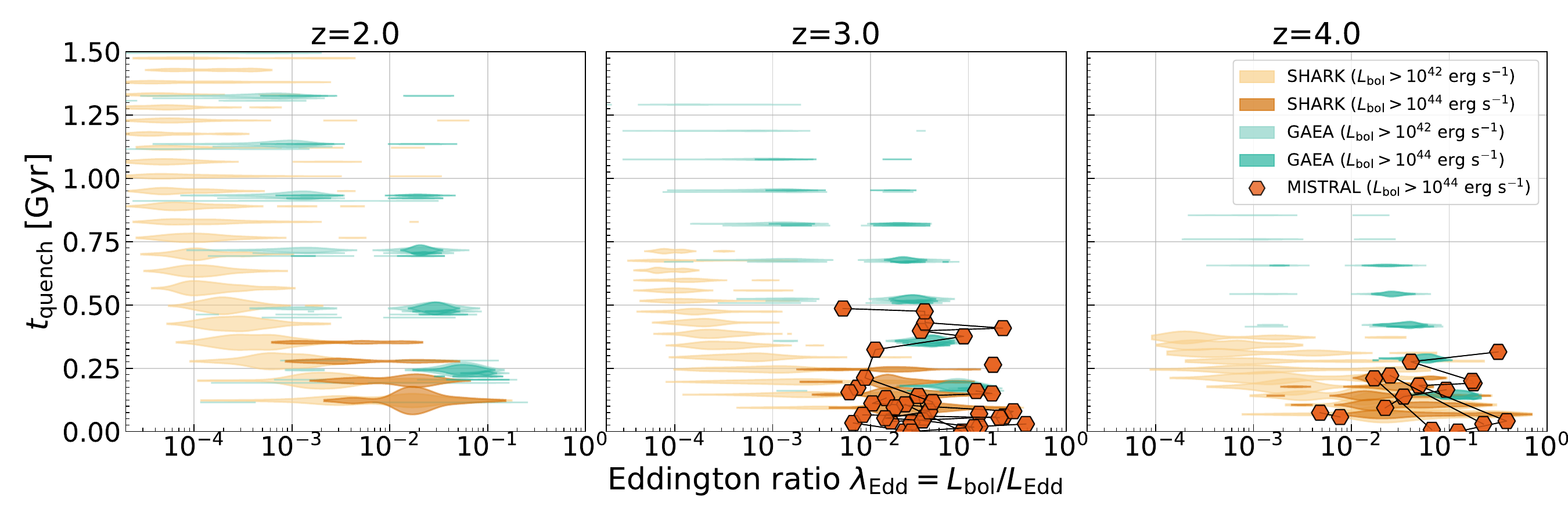}
    \caption{Eddington ratio $\lambda_{\rm Edd} = L_{\rm bol}/L_{\rm Edd}$ versus time since quenching $t_{\rm quench}$ for galaxies in the \shark\ (orange) and \gaea\ (teal) SAMs  at $z=2$, $3$, and $4$. For each model, two AGN luminosity thresholds are shown: $L_{\rm bol}>10^{42}$ \es\ (light shading) and $L_{\rm bol}>10^{44}$ \es\ (dark shading). At each discrete value of $t_{\rm quench}$, the horizontal extent of each patch traces the KDE of the \eddratio\ distribution, with the patch width proportional to the number of galaxies at each time step. Individual orange hexagons show \blackdawn/\mistral\ zoom-in results, with connecting lines linking AGN episodes within the same galaxy.}
    \label{fig:edd_ratio_tquench}
\end{figure*}
Figure \ref{fig:model_distributions} shows the distributions of \mstar, \mbh, their ratio \mbh/\mstar, and \eddratio\ at $z=2-4$ for simulated quiescent galaxies with AGN activity, compared with our \nev- and \oiii-detected samples. For the \blackdawn\ simulations, which stop at $z=3$, we treat each recorded nuclear activity event of each quenched galaxy as a single measurement, giving a total of 54 measurements for 18 galaxies, with masses estimated in snapshots between $z=3-4.4$.\footnote{Multiple AGN events recorded over time for a single object are not completely independent, as they occur within the same galaxy.}

All models produce sources with \lbol\ in excess of $10^{44}$ \es, consistent with the lower end of our observed values. Notably, they predict AGN with \eddratio\ as high as $10$--$50$\%, particularly at the highest redshifts and youngest ages immediately following quenching, consistent with the values inferred from the observations. \shark\ produces fewer massive quenched galaxies at high redshift than in the {\sc Black Dawn} simulations, partly because the parent sample of the latter is drawn from the most massive galaxies in \tng-100, and partly because of the details of the AGN feedback model. Indeed, the stellar masses in \blackdawn\ are lower than in the \tng-100 parent sample, as the \mistral\ feedback model suppresses star formation earlier than the standard TNG physics does \citep{farcy_2025}. \gaea\ produces stellar masses intermediate between those of \shark\ and \blackdawn. As with the \mstar\ distributions, the \mbh\ values produced by \shark\ are slightly lower than those in \gaea\ and \blackdawn\ at $z=3$. At higher redshift, \blackdawn\ produces lower \mbh\ values than the two SAMs. This is reflected in the normalization of the \mbh--\mstar\ relation, which is similar for \gaea\ and \shark\ (though at different absolute values of \mstar\ and \mbh), while the \mistral\ recipe tends to produce less massive black holes per unit stellar mass. All are formally consistent with the local \mbh--\mstar\ relation for massive galaxies within its large scatter ($M_{\rm BH}/M_\star \sim 0.001-0.01$, \citealt{kormendy_2013, greene_2020, bravo_2025}). The Eddington ratios follow directly from the trends in \lbol\ and \mbh.

Beyond absolute values, the models broadly reproduce a key trend: the strongest activity (in terms of \lbol) is preferentially found among the youngest quenched galaxies, while older systems do not typically host strong, radiatively efficient QSOs. This is reproduced by both \shark\ and \gaea, corresponding to the buildup of the population dominated by old stellar populations from $z=4$ to $z=2$ and an overall decrease in \lbol\ and \eddratio\ with time, consistent with the well-established decline of QSO luminosities toward lower redshifts. 
This is clear from Figure \ref{fig:edd_ratio_tquench}, where we show the distribution of \eddratio\ at each discrete value of $t_{\rm quench}$ for QGs with AGN activity above two \lbol\ thresholds for \shark\ and \gaea. As noted above, selecting galaxies above higher \lbol\ thresholds naturally skews the sample toward more recently quenched galaxies at all redshift snapshots, while the overall distribution of QGs is progressively dominated by less active systems. As cosmic time progresses, galaxies with older $t_{\rm quench}$ appear and their \lbol\ and \eddratio\ are generally low. In absolute terms, the evolution of the predicted distributions differs between \shark\ and \gaea, a distinction that larger observational samples could test. We note that the location and width of the gap in the distributions of \eddratio\ at fixed $t_{\rm quench}$ are due to the implementation of black hole accretion and feedback in both models. Results for individual \blackdawn\ snapshots of QGs with strong AGN activity ($L_{\rm bol}>10^{44}$ \es) also align with the tails predicted by \shark\ and \gaea. Related results are reported by \citet{barchiesi_2025}, who matched their sample of Type 2 obscured X-ray AGN hosts with \nev\ at $z\sim1$ to simulated galaxies from {\sc Simba} \citep{dave_2019}; the main difference with our findings is that their \nev\ emitters are mostly in a transient ``pre-quenching'' phase, with an SFR not significantly different from its peak value, a population we cannot capture since our selection requires quenched systems in the first place.
 
It is interesting to note that rejuvenation events due to gas-rich mergers or gas reaccretion via fountains after quenching \citep{zhu_2026_outflows, taylor_2026} can reactivate substantial nuclear activity while only mildly affecting the total \mstar\ via small bursts of star formation. In the runs with the \mistral\ feedback model, galaxies in \blackdawn\ typically quench $\sim10$ Myr after the Eddington ratio of their black hole drops below $10$\%, corresponding to the median time needed for the AGN wind to propagate beyond the black hole's surroundings and suppress star formation, with black hole accretion affected before star formation. Sustained AGN activity ($\lambda_{\rm Edd} > 0.1$) reappears during the quenching phase in $\sim20$\% of episodes after a median delay of 57 Myr. When considering AGN activity both during and after quenching, including episodes with rejuvenation and gas reaccretion, this fraction rises to $43$\% with a median delay of 122 Myr, consistent with our back-of-the-envelope estimate. Rejuvenation is also a significant process in \gaea\ \citep{delucia_2024,fontanot_2025b} and other models at high redshift \citep[e.g.,][]{remus_2025_magneticum}, while becoming less important with cosmic time \citep{chandro-gomez_2025}.

Taken together, the models analyzed here consistently show that QSO-like AGN activity can persist in galaxies after quenching at high redshift before fading as stellar populations age, underscoring the importance of modeling very high accretion rates to reproduce the earliest quenched galaxies. Current models span the parameter space occupied by our observations, yet discriminating between feedback recipes will require systematic follow-up tracing the full distributions of observed galaxy properties. Searching for high-ionization lines such as \nev\ provides a clean and powerful means of identifying the brightest and most extreme instances of post-quenching AGN activity, and assembling the statistical samples needed to do so.

\section{Conclusions}

We report the detection of the high-ionization line \nev$\lambda$3427 in the \textit{JWST}/NIRSpec spectra of 6 massive quenched galaxies at $z\sim1.5-4.5$, selected from an archival sample of 87 systems with spectral coverage of the line. This galaxy-first approach---selecting bona fide quenched systems and subsequently searching for signatures of ongoing SMBH accretion---provides a complementary, clean probe of the brightest, most radiatively efficient AGN activity after quenching in the distant Universe, free from contamination by residual star formation. Our main findings are as follows:

\begin{list}{$\bullet$}{\leftmargin=1.5em}

    \item All \nev-detected sources occupy the AGN locus of the BPT diagram and display high \oiii/\hb\ and \nii/\ha\ ratios, confirming that \nev\ reliably selects the most extreme tail of AGN activity in distant quenched galaxies. The brightest X-ray detections are also detected in \nev, with $L_{\rm 2-10\,keV}/L_{\rm [NeV]} \approx 300-1000$, and show broad-line \ha\ profiles, while the remaining \nev\ detections are undetected in X-ray, with upper limits consistent with the obscured AGN population studied at lower redshift.

    \item Empirical \mbh\ measurements from broad \ha\ emission ($\mathrm{FWHM} \gtrsim 4000$ \kms) place our \nev-detected QGs on the local \mbh--\mstar\ scaling relation. The bolometric luminosities inferred from \nev\ and \oiii\ reach quasar-like values ($L_{\rm bol} = 10^{45-46}$ \es), with Eddington ratios of $\lambda_{\rm Edd} \approx 10-50$\% and black hole accretion rates of a few \myr, comparable to or exceeding the residual star formation rates in the most extreme cases.

    \item The strongest \nev\ emitters are preferentially found among the most recently quenched systems with $D_n{4000} \lesssim 1.3$, while the high-\dn, high-\eddratio\ region of parameter space is systematically empty, demonstrating that radiatively efficient SMBH growth via gas accretion can persist several hundred Myr after the main quenching epoch before fading as stellar populations age. The estimated AGN duty cycle, based on the \nev\ detection fraction, is of the order of $200$ Myr, decreasing to $66^{+83}_{-21}$ Myr for the most luminous sources with $L_{\rm bol} > 10^{46}$ \es, consistent with estimates for QSOs at similar redshifts.

    \item Semi-analytical models (\shark, \gaea) and zoom-in simulations ({\sc Black Dawn} with the \mistral\ feedback model) broadly reproduce the observed trend of declining AGN activity with time since quenching, consistent with the physical picture emerging from the observations. At peak post-quenching activity, all models produce Eddington ratios of $\approx 10-50$\% and bolometric luminosities broadly consistent with our observed values. 

\end{list}

Systematic follow-up with \textit{JWST}/NIRSpec targeting \nev\ in a mass-complete sample of quenched galaxies will simultaneously deliver, at no additional cost, the full rest-frame optical continuum and stellar absorption features redward of the Balmer break that neither ground-based facilities nor other space observatories can reach with sufficient sensitivity at high redshift. This will enable accurate measurements of AGN duty cycles and a comprehensive mapping of the transition from QSO-dominated to stellar-continuum-dominated spectra across cosmic time.

\begin{acknowledgements}
 We warmly thank Luigi Barchiesi for providing data for his sample of obscured AGN. 
 FV, KI, and PZ acknowledge support from the Independent Research Fund Denmark (DFF) under grant 3120-00043B. PA-A acknowledges support from DFF under grant 4251-00086B. MO acknowledges support by JSPS KAKENHI Grant Number JP25K07361. 
 TK acknowledges support from JSPS grant 25KJ1331.
 This work is based in part on observations made with the NASA/ESA/CSA James Webb Space Telescope. The data were obtained from the Mikulski Archive for Space Telescopes at the Space Telescope Science Institute, which is operated by the Association of Universities for Research in Astronomy, Inc., under NASA contract NAS 5-03127 for JWST. MH and KEW acknowledge support for program JWST-GO-3567, provided by NASA through a grant from the Space Telescope Science Institute, which is operated by the Association of Universities for Research in Astronomy, Inc., under NASA contract NAS 5-03127. Some of the data products presented herein were retrieved from the Dawn JWST Archive (DJA). DJA is an initiative of the Cosmic Dawn Center, which is funded by the Danish National Research Foundation under grant DNRF140.
\end{acknowledgements}

\bibliographystyle{aa} 
\bibliography{bib_nev}

\begin{appendix}
\section{Data table}

\begin{landscape}
\begin{table}
\centering
\caption{Measurements and physical properties of the \nev\ detected sample.}
\label{tab:table}
\begin{tabular}{ccccccccccccc}
\toprule
\toprule
ID & Archive & RA & Dec & $z_{\rm spec}$ & $\sigma_\star$ & $\sigma_{\rm gas}$ & $\sigma_{\rm broad}$ & $L_{\rm [NeV]3427}$ & $L_{\rm [OII]}$ & $L_{\rm [NeIII]3869}$ & $L_{\rm H\beta}$ & $L_{\rm [OIII]5007}$ \\
 &  & \small [deg] & \small [deg] & & \small [\kms] & \small [\kms] & \small [\kms] & \small [$10^{41}$ \es] & \small [$10^{41}$ \es] & \small [$10^{41}$ \es] & \small [$10^{41}$ \es] & \small [$10^{42}$ \es] \\
\midrule
 78 & DD  & 34.298697  &  $-4.989901$ & $4.0110(3)$ & $240^{+25}_{-24}$  & $446^{+32}_{-29}$  & $3237^{+231}_{-214}$ & $8.60^{+1.62}_{-1.69}$ & $8.48^{+1.26}_{-1.46}$ & $12.24^{+1.64}_{-1.62}$ & $1.93^{+1.40}_{-1.48}$ & $3.45^{+0.19}_{-0.19}$ \\
 82 & DD  & 34.316190  &  $-5.051441$ & $3.9832(7)$ & $178^{+54}_{-79}$  & $308^{+3}_{-3}$    & $\dots$              & $3.82^{+0.35}_{-0.35}$ & $8.01^{+0.32}_{-0.32}$ & $3.73^{+0.38}_{-0.35}$ & $5.95^{+0.27}_{-0.28}$ & $5.32^{+0.05}_{-0.05}$ \\
111 & DD  & 34.289452  &  $-5.269803$ & $3.7968(3)$ & $208^{+28}_{-29}$  & $198^{+7}_{-7}$    & $4265^{+195}_{-199}$ & $1.46^{+0.43}_{-0.43}$ & $3.37^{+0.43}_{-0.46}$ & $1.63^{+0.37}_{-0.39}$ & $1.58^{+0.33}_{-0.35}$ & $1.95^{+0.04}_{-0.04}$ \\
329 & DJA & 53.165314  & $-27.814140$ & $3.0640(3)$ & $335^{+25}_{-28}$  & $440^{+19}_{-19}$  & $8308^{+741}_{-846}$ & $1.86^{+0.66}_{-0.68}$ & $7.32^{+0.78}_{-0.85}$ & $3.39^{+1.29}_{-1.13}$ & $4.84^{+0.70}_{-0.71}$ & $1.46^{+0.35}_{-0.37}$ \\
924 & DJA & 150.133502 &   2.370408 & $2.0074(3)$ & $174^{+35}_{-35}$  & $223^{+7}_{-7}$    & $\dots$              & $1.15^{+0.31}_{-0.33}$ & $2.39^{+0.26}_{-0.24}$ & $0.69^{+0.20}_{-0.21}$ & $1.59^{+0.22}_{-0.25}$ & $1.01^{+0.04}_{-0.04}$ \\
Ito+25& $-$ & 150.074583 & 2.302000 & $2.0943(3)$& $238^{+32}_{-32}$& $238_{-6}^{+6}$& $1854^{+34}_{-34}$& $0.85^{+0.11}_{-0.11}$& $1.09_{-0.34}^{+0.34}$& $0.78^{+0.11}_{-0.11}$& $1.70^{+0.19}_{-0.19}$& $0.95^{+0.02}_{-0.02}$\\
\bottomrule
\vspace*{0.2cm}
\end{tabular}
\begin{tabular}{ccccccccccccc}
\toprule
\toprule
ID & $L_{\rm H\alpha}$ & $L_{\rm [NII]6583}$ & $L_{\rm H\beta,broad}$ & $L_{\rm H\alpha,broad}$ & $L_{\rm bol,[NeV]}$ & $L_{\rm bol,[OIII]}$ & $\rm BHAR_{\rm [NeV]}$ & $\rm BHAR_{\rm [OIII]}$ & $\log M_{\rm BH}$ & $\log L_{\rm Edd}$ & $\lambda_{\rm Edd,[NeV]}$ & $\lambda_{\rm Edd,[OIII]}$ \\
 & \small [$10^{42}$ \es] & \small [$10^{42}$ \es] & \small [$10^{41}$ \es] & \small [$10^{42}$ \es] & \small [$10^{46}$ \es] & \small [$10^{46}$ \es] & \small [\myr] & \small [\myr] & \small [\msun] & \small [\es] & & \\
\midrule
 78 & $0.80^{+0.23}_{-0.22}$  & $4.18^{+0.21}_{-0.26}$ & $19.77^{+3.97}_{-4.54}$ & $7.05^{+0.56}_{-0.63}$ & $3.84^{+0.73}_{-0.75}$ & $1.20^{+0.07}_{-0.07}$ & $6.10^{+1.15}_{-1.20}$ & $1.90^{+0.10}_{-0.11}$ & $8.82^{+0.07}_{-0.07}$ & $46.92^{+0.07}_{-0.07}$ & $0.46^{+0.11}_{-0.10}$ & $0.15^{+0.03}_{-0.02}$ \\
 82 & $2.23^{+0.04}_{-0.04}$  & $2.20^{+0.04}_{-0.04}$ & $\dots$                 & $\dots$                 & $1.71^{+0.15}_{-0.16}$ & $1.85^{+0.02}_{-0.02}$ & $2.71^{+0.24}_{-0.25}$ & $2.93^{+0.03}_{-0.03}$ & $\dots$                & $\dots$                 & $\dots$                & $\dots$                \\
111 & $0.75^{+0.03}_{-0.03}$  & $0.91^{+0.03}_{-0.03}$ & $0.25^{+1.76}_{-0.25}$  & $2.70^{+0.16}_{-0.15}$ & $0.65^{+0.19}_{-0.19}$ & $0.67^{+0.01}_{-0.02}$ & $1.04^{+0.31}_{-0.31}$ & $1.07^{+0.02}_{-0.02}$ & $8.87^{+0.05}_{-0.05}$ & $46.97^{+0.05}_{-0.05}$ & $0.07^{+0.02}_{-0.02}$ & $0.07^{+0.01}_{-0.01}$ \\
329 & $1.61^{+0.23}_{-0.24}$  & $5.70^{+0.39}_{-0.37}$ & $\dots$                 & $2.83^{+0.56}_{-0.50}$ & $0.83^{+0.29}_{-0.30}$ & $0.51^{+0.12}_{-0.13}$ & $1.32^{+0.47}_{-0.48}$ & $0.80^{+0.19}_{-0.21}$ & $9.47^{+0.10}_{-0.12}$ & $47.57^{+0.10}_{-0.12}$ & $0.02^{+0.01}_{-0.01}$ & $0.013^{+0.006}_{-0.004}$ \\
924 & $1.06^{+0.04}_{-0.04}$  & $1.39^{+0.04}_{-0.04}$ & $\dots$                 & $\dots$                 & $0.51^{+0.14}_{-0.15}$ & $0.35^{+0.01}_{-0.01}$ & $0.81^{+0.22}_{-0.23}$ & $0.56^{+0.02}_{-0.02}$ & $\dots$                & $\dots$                 & $\dots$                & $\dots$                \\
Ito+25& $0.30^{+0.05}_{-0.05}$& $0.27^{+0.04}_{-0.04}$& $14.0^{+0.4}_{-0.4}$& $12.48^{+0.16}_{-0.16}$ & $0.38^{+0.05}_{-0.05}$& $0.33^{+0.01}_{-0.01}$& $0.60^{+0.08}_{-0.08}$& $0.53^{+0.01}_{-0.01}$& $8.43^{+0.02}_{-0.02}$& $46.53^{+0.02}_{-0.02}$& $0.11^{+0.01}_{-0.01}$& $0.097^{+0.002}_{-0.002}$\\
\bottomrule
\vspace*{0.2cm}
\end{tabular}
\begin{tabular}{ccccccccc}
\toprule
\toprule
ID & $L_{\rm [NeV]}/L_{\rm [NeIII]}$ & $L_{\rm [OIII]}/L_{\rm H\beta}$ & $L_{\rm [NII]}/L_{\rm H\alpha}$ & $M_\star$ & SFR & $D_n4000$ & X-ray flag & $L_{\rm2-10\,keV}$ \\
 & & & & \small [$10^{10}$ \msun] & \small [\myr] & & & \small [$10^{43}$ \es] \\
\midrule
 78  & $0.70^{+0.18}_{-0.17}$ & $17.95^{+57.83}_{-7.71}$ & $5.19^{+2.27}_{-1.27}$ & $11.71 \pm 1.07$   & $<22.8$   & $1.102 \pm 0.014$ & 1 & $25.8$   \\
 82  & $1.02^{+0.16}_{-0.12}$ & $8.93^{+0.46}_{-0.40}$   & $0.99^{+0.03}_{-0.03}$ & $2.25 \pm 0.32$  & $21.49 \pm 11.66$ & $1.041 \pm 0.045$ & 0 & $<2.97$  \\
111  & $0.89^{+0.39}_{-0.29}$ & $12.31^{+3.45}_{-2.10}$  & $1.21^{+0.06}_{-0.06}$ & $6.95 \pm 0.74$  & $<63.3$  & $1.148 \pm 0.018$ & 1 & $10.4$   \\
329  & $0.55^{+0.34}_{-0.23}$ & $3.01^{+0.83}_{-0.77}$   & $3.55^{+0.50}_{-0.37}$ & $12.77 \pm 0.91$  & $<8.7$   & $1.173 \pm 0.019$ & 1 & $26.6$   \\
924  & $1.68^{+0.92}_{-0.60}$ & $6.35^{+1.24}_{-0.76}$   & $1.32^{+0.06}_{-0.06}$ & $12.28 \pm 1.64$   & $<47.7$  & $1.433 \pm 0.029$ & 0 & $<1.27$  \\
Ito+25& $1.09_{-0.19}^{+0.19}$& $5.62^{+0.65}_{-0.65}$& $0.93^{+0.15}_{-0.15}$& $16\pm2$& $<12.8$& $1.304\pm0.020$& 1 & $8.9$\\
\bottomrule
\end{tabular}
\tablefoot{IDs as in \cite{ito_2025_deepdive}; Archive: DeepDive (DD), Dawn JWST Archive (DJA); $\sigma_\star$, $\sigma_{\rm gas}$, and $\sigma_{\rm broad}$ are the velocity dispersions of the stellar, narrow gas, and broad gas components, respectively; $L_{\rm line}$ is the luminosity of each line in the narrow gas component as labeled without dust correction; $L_{\rm line,\,broad}$ is the luminosity of each line in the broad component without dust correction; $L_{\rm bol, [NeV]}$ and $L_{\rm bol, [OIII]}$ are the AGN bolometric luminosities derived from \nev\ and \oiii, respectively; BHAR is the derived black hole accretion rate; the black hole mass \mbh\ is derived from the broad \ha\ component as in \cite{reines_2013}; $L_{\rm Edd}$ and \eddratio\ are the ensuing Eddington luminosities and ratios; \mstar, SFR, \dn, X-ray flag (1=detection, 0=upper limit), and $L_{\rm 2-10\,keV}$ are from \cite{ito_2025_deepdive} and reported here for convenience. Upper limits are at $3\sigma$ significance. A machine-readable version of this table, including the rest of the sample with \nev\ coverage, but not detected, is available online.}
\end{table}
\end{landscape}
\end{appendix}

\end{document}